\documentclass[final,5p,times,twocolumn]{elsarticle}

\usepackage[utf8x]{inputenc}
\usepackage[cmex10]{amsmath}
\usepackage{amsfonts}
\usepackage{amsthm}
\usepackage{algorithmic}

\usepackage{amsfonts}
\usepackage{amssymb}
\usepackage{graphicx}
\usepackage{url}
\usepackage{hyperref}
\usepackage{color}

\begin{document}
\begin{frontmatter}
%
\title{Wave-absorbing vehicular platoon controller}

\author[label1]{Dan~Martinec,
        Ivo~Herman,
        Zden\v{e}k~Hur\'{a}k,
        and~Michael~\v{S}ebek}

\address[label1]{All authors are with the Department of Control Engineering, Czech Technical University in Prague, Prague, Czech Republic, (e-mails: martinec.dan@fel.cvut.cz; ivo.herman@fel.cvut.cz; hurak@fel.cvut.cz sebekm1@fel.cvut.cz)}


\begin{abstract}
The paper tailors the so-called wave-based control popular in the field of flexible mechanical structures to the field of distributed control of vehicular platoons. The proposed solution augments the symmetric bidirectional control algorithm with a wave-absorbing controller implemented on the leader, and/or on the rear-end vehicle. The wave-absorbing controller actively absorbs an incoming wave of positional changes in the platoon and thus prevents oscillations of inter-vehicle distances. The proposed controller significantly improves the performance of platoon manoeuvrers such as acceleration/deceleration or changing the distances between vehicles without making the platoon string unstable. Numerical simulations show that the wave-absorbing controller performs efficiently even for platoons with a large number of vehicles, for which other platooning algorithms are inefficient or require wireless communication between vehicles.
\end{abstract}


\end{frontmatter}


\section{INTRODUCTION}
\subsection{Vehicular platooning}
The field of \emph{vehicular platooning} was active as early as the 1960's and remains so until now. The task is to safely and effectively control several vehicles driving behind each other, for example on a highway lane. It is motivated by higher throughput, lower fuel consumption, increase of traffic safety etc.

Regarding control strategies; among the first treatments of vehicular platooning were papers by \cite{Levine1966} and \cite{Melzer1971}. They examined a centralized control approach with a single global controller governing all vehicles. However, \cite{Jovanovic2005a} later showed that one has to be careful about the stabilizability of the system, since it might degrade with increasing number of vehicles. Nevertheless, more attention is paid to fully or partially distributed control, wherein each vehicle is controlled by its own on-board controller with only limited knowledge about the platoon. Among the first papers dealing with the distributed control was work by \cite{chu_decentralized_1974}. Basic questions about the feasibility and performance of such systems was introduced by \cite{Cosgriff1969} and later formalized by \cite{Swaroop1996} under the term \emph{string stability}. String stability, or more precisely string instability, is a~phenomenon that causes higher control demands on the members of a~vehicular platoon that are further from the source of regulation error. Although string stability does not guarantee that the vehicles do not crash into each other, it is a useful analysis tool. A way how a regulation error or a disturbance propagates in a platoon of vehicles controlled by various distributed control strategies was examined in several papers, see for instance \cite{Seiler2004a}, \cite{Barooah2005a} and \cite{Shaw2007}. A fundamental limitation of many distributed algorithms with only local information about the platoon is inability to maintain coherence in a~large-scale platoon subjected to stochastic disturbances \cite{Bamieh2012b}. Though, the coherence can be improved by introducing optimal non-symmetric localized feedback \cite{Lin2011}.

A common goal of each platooning algorithm is to drive the platoon with a reference velocity and inter-vehicle distances. Many distributed algorithms have been introduced in the platooning field. The most simple algorithm relying only on the measurement of the distance to the immediately preceding vehicle is the so-called \emph{predecessor following algorithm}. A straightforward extension is the so-called \emph{bidirectional control algorithm}, which additionally measures the distance to the immediate follower. Depending on the weight between these two distance measurements, we distinguish either \emph{symmetric} or \emph{asymmetric} bidirectional control. Although, the asymmetric version improves the stability in terms of the least stable closed-loop eigenvalue as proved by \cite{Barooah2009}, we let our in-platoon vehicles to be controlled by the symmetric version, analysed for instance in \cite{Lestas2007} or \cite{Middleton2010}. The reason for doing this will be clear after Section~\ref{sec:reflections}.

\subsection{Wave-based control concept}
The origins of the control based on travelling waves lies in the 1960's in mathematical modeling and analysis of flexible structures. Paper of \cite{vaughan_application_1968} was one of the first treatments analysing simpler instances of flexible structures such as beams and plates. Analysis and control of a more complex flexible structures from the viewpoint of travelling wave-modes was investigated in a series of papers by von Flotow and his colleagues in \cite{Flotow1985} and \cite{Flotow1986}.

Recently, the concept was revisited in a series of papers by O'Connor in \cite{OConnor2006} and \cite{OConnor2007} for vibrationless positioning of lumped multi-link flexible mechanical systems. It was named \emph{wave-based control} and it is based on the so-called \emph{wave transfer function}, which describes how the traveling wave propagates in the lumped system. Simultaneously with O'Connor, the wave concept was also revisited for a control of continuous flexible structures by \cite{Halevi2005} under the name \emph{absolute vibration suppression}. It relies on the transfer function as well, though in this case, the time delay plays a key role. Surprisingly, it was shown by the joint paper of the last two mentioned authors in \cite{Peled2012}, that both the wave-based control and the absolute vibration suppression are just a feedback version of the input shaping control. It was also shown that the wave-based control can be generalized even for continuous flexible systems, e.g. a steel rod, and then it coincides with the absolute vibration suppression.

The key idea of the wave-based control is to generate a wave at the actuated front end of the interconnected system and let it propagate to the opposite end of the system, where it reflects and returns back to the front-end actuator. When it reaches the front again, it is absorbed by the front-end actuator by means of the wave transfer function. A both interesting and troublesome property of the wave transfer function is the presence of the square root of polynomial in the function. This makes its implementation in the time domain very challenging. To be able to run numerical simulations, we therefore introduce a convergent recursive algorithm that approximates the wave transfer function for an arbitrary dynamics of the local system.

There are other viewpoints on the wave-based control. One was introduced by \cite{Ojima2001} in terms of the characteristic impedance for a mass-spring system. Other possible viewpoint introduced \cite{Nagase2005} for wave control of ladder electric networks.

\subsection{Objective of the paper}
In this paper, a finite one-dimensional platoon of vehicles moving in a highway lane is considered. Each individual vehicle in the platoon is locally controlled by a bidirectional controller, which plays the role of string-damper connection in mechanical structures and hence enables a wave to propagate back and forth. One or both of the platoon ends are controlled by the \emph{wave-absorbing controller} allowing active absorption of the traveling wave. The similarity of bidirectional control with continuous wave equation was described in \cite{Herman2013}

The key objective of the paper is to generalize the principle of the wave-based control used in the field of mechanics for vehicular platooning control in such a way that the distances between vehicles are additionally considered. In this regard, the presented concept offers a symmetric version of bidirectional control enhanced by the feedback control of one or both platoon ends. Thus, it significantly decreases long transient oscillations during platoon manoeuvres such as acceleration/deceleration or changing the distances between vehicles. In addition, the paper contributes the following: a) It generalizes the wave transfer function description for the arbitrary dynamics of the local system, b) it offers the convergent recursive algorithm that approximates the wave transfer function, c) it presents an alternative way of deriving the wave transfer function using a continued fraction approach, and d) it provides a mathematical derivation of the transfer functions describing reflections on the platoon ends.

The paper is structured as follows. Section 2 gives a mathematical model of the vehicle. Section 3 describes the wave transfer function as a requisite tool for the wave description. A mathematical description of wave reflections on forced and free ends is given in Section 4. Section 5 introduces the wave-absorbing controller as an addon for the bidirectional control. The new controllers are analyzed by numerical simulations in section 6. The necessary mathematical derivations are given in the three appendices.

\section{LOCAL CONTROL OF THE PLATOON VEHICLES}
A vehicle in a platoon indexed by $n$ is in the Laplace domain modelled as
\begin{align}
  X_{n}(s) = P(s) U_n(s),
  \label{eq:system_car}
\end{align}
where $s$ is the Laplace variable, $X_n(s)$ is a position of the $n$th vehicle in the Laplace domain, $P(s)$ represents the transfer function of system dynamics and $U_n(s)$ is the system input which is generated by the local controller of the vehicle specified in the following.

Except for the leader indexed $n=0$ and the rear-end vehicle, each vehicle in the platoon is equipped with a symmetric bidirectional controller $C(s)$ with the task of equalizing the distances to its immediate predecessor and successor, giving
\begin{align}
U_n(s) = C(s)(D_{n-1}(s) - D_{n}(s)),
\label{eq:orig_diff_eq_long}
\end{align}
where $D_n(s)$ is the distance between vehicles indexed by $n$ and $n+1$, hence $D_n(s)= X_{n}(s) - X_{n+1}(s)$. Substituting (\ref{eq:orig_diff_eq_long}) into (\ref{eq:system_car}) yields the resulting model of the in-platoon vehicle with the bidirectional control for the inter-vehicle distances,
\begin{align}
X_{n}(s) = P(s)C(s) (X_{n-1}(s)-2X_{n}(s)+X_{n+1}(s)).
\label{eq:orig_diff_eq_simp}
\end{align}
Using the notation,
\begin{equation}
  \alpha(s) = \frac{1}{P(s)C(s)}+2,
  \label{eq:alpha_generalized}
\end{equation}
equation (\ref{eq:orig_diff_eq_simp}) is thus rewritten as
\begin{align}
  X_{n}(s) &= \frac{1}{\alpha(s)}(X_{n-1}+X_{n+1}).
  \label{eq:bidir_sys_laplace}
\end{align}
The vehicle at the rear end of the platoon is driven by the predecessor following algorithm and is supposed to equalize the distance to its immediate predecessor and reference distance $D_{\text{ref}}$,
\begin{align}
  X_{N}(s) = \frac{1}{\alpha(s)-1}(X_{N-1}(s)-D_{\text{ref}}(s)),
  \label{eq:last_sys_laplace}
\end{align}
where $X_{N}(s)$ is the position of the last vehicle in the platoon.

To carry out numerical simulations, we will use the model that is often used in theoretical studies. The vehicle is described by a double integrator model with a simple (linear) model of friction, $\xi$, and controlled by a PI controller. Hence, $P(s) = 1/(s^2+\xi s)$ and $C(s) = (k_{\text{p}} s +k_{\text{i}})(s)$, where $k_{\text{p}}$ and $k_{\text{i}}$ are proportional and integral gains of the PI controller, respectively. Such a model was also used in experimental studies in \cite{Martinec2012}.

\section{WAVE TRANSFER FUNCTION}

The bidirectional property of locally controlled systems causes any change in the movement of the leading vehicle to propagate through the platoon as a \emph{wave} up to the last vehicle. To describe this wave, we need to find out how the position of a vehicle is influenced by the position of its immediate neighbours. For a moment, let us assume that the length of the platoon is infinite, so that there is no platoon end where the wave can reflect. A generalization for platoon with a one platoon end, i.e. a semi-infinite platoon, is done in the next section.

\subsection{Mathematical model of the wave transfer function}
Following the standard arguments for \emph{wave equation} found for instance in \cite{Asmar2004}, the solution to the wave equation can be decomposed into two components: $A_{n}(s)$ and $B_n(s)$ (also called \emph{wave variables} in the literature), which represent two waves propagating along a platoon in the forward and backward directions, respectively.

To find a transfer function describing the wave propagation, we are searching for two linearly independent recurrence relations that satisfy (\ref{eq:bidir_sys_laplace}). We first recursively apply (\ref{eq:bidir_sys_laplace}) and (\ref{eq:last_sys_laplace}) with $D_{\text{ref}}(s) = 0$, for a platoon with an increasing number of vehicles. The transfer function for a platoon with two vehicles is $A_1 / A_0 = (\alpha-1)^{-1}$, for a platoon with three vehicles is $A_1/A_0 = \left(\alpha-(\alpha-1)^{-1} \right)^{-1}$, for a platoon with four vehicles is $A_1/A_0 = \left(\alpha - \left( \alpha-(\alpha-1)^{-1} \right)^{-1} \right)^{-1}$ and so on. Continuing recursively, $A_1/A_0$ is expressed by the continued fraction
\begin{equation}
  \dfrac{A_1}{A_0} = \dfrac{1}{\alpha-\dfrac{1}{\alpha-\dfrac{1}{\alpha-\dfrac{1}{\ddots}}}}.
  \label{eq:trans_continued_fraction}
\end{equation}
The continued-fraction expansion of a square root is given by \cite{jones1984}
\begin{equation}
  \sqrt{z^2+y} = z+\dfrac{y}{2z+\dfrac{y}{2z+\dfrac{y}{2z+\dfrac{y}{\ddots}}}}.
  \label{eq:sqrt_continued_fraction}
\end{equation}
Letting the number of vehicles approach infinity, the right-hand sides of (\ref{eq:trans_continued_fraction}) and (\ref{eq:sqrt_continued_fraction}) are equal, provided that $y=-1$ and $z=\alpha/2$. Hence,
\begin{equation}
  \frac{A_{1}}{A_0} = \frac{\alpha}{2}-\frac{1}{2}\sqrt{\alpha^2-4}.
  \label{eq:G1_continued_fraction}
\end{equation}
Likewise, the transfer function $A_2/A_1$ can be expressed from (\ref{eq:bidir_sys_laplace}) and (\ref{eq:last_sys_laplace}) for $n = 2$ as
\begin{align}
  \alpha A_1 &= A_0+A_2.
\end{align}
Substituting for $A_0$ from the previous recursive step (\ref{eq:G1_continued_fraction}) gives
\begin{align}
  \alpha A_1 &= A_1\left(\frac{\alpha}{2}+\frac{1}{2}\sqrt{\alpha^2-4}\right) + A_2,
\end{align}
which provides
\begin{align}
  \frac{A_2}{A_1} &= \frac{\alpha}{2}-\frac{1}{2}\sqrt{\alpha^2-4}.
  \label{eq:G1_continued_fraction2}
\end{align}
Continuing recursively, we can find that the transfer function $A_{n+1}/A_{n}$ is again equal to (\ref{eq:G1_continued_fraction}) or (\ref{eq:G1_continued_fraction2}). We can conclude that the transfer function from the $n$th to $(n+1)$th vehicle is the same for each vehicle, and is equal to
\begin{equation}
  G_1(s) = \frac{\alpha}{2}-\frac{1}{2}\sqrt{\alpha^2-4}. \label{eq:G1_WTF}
\end{equation}

Analogously, the second linearly independent recurrence relation of (\ref{eq:bidir_sys_laplace}) and (\ref{eq:last_sys_laplace}) is searched for by their recursive application with a decreasing index of vehicles. After similar algebraic manipulations as for $A_n$, we find
\begin{equation}
  \dfrac{B_n}{B_{n-1}} = \alpha- \dfrac{1}{\alpha-\dfrac{1}{\alpha-\dfrac{1}{\alpha-\dfrac{1}{\ddots}}}}.
  \label{eq:trans_continued_fraction_B}
\end{equation}
Letting the number of vehicles approach infinity, the right-hand sides of (\ref{eq:trans_continued_fraction_B}) and (\ref{eq:sqrt_continued_fraction}) are equal provided that $y=-1$ and $z=\alpha/2$. Hence,
\begin{equation}
  \frac{B_n}{B_{n-1}} = \frac{\alpha}{2}+\frac{1}{2}\sqrt{\alpha^2-4}.
  \label{eq:G2_continued_fraction}
\end{equation}
The transfer function from $n$th to $(n-1)$th vehicle is the same for each vehicle, and is equal to
\begin{equation}
  G_2(s) = \frac{\alpha}{2}+\frac{1}{2}\sqrt{\alpha^2-4}. \label{eq:G2_WTF}
\end{equation}

The resulting model of the vehicular platoon with an infinite number of vehicles is therefore described as follows:
\begin{align}
  X_n &= A_n + B_n,
  \label{eq:pos_decomp}\\
  A_{n+1} &= G_1A_n,
  \label{eq:anp1}\\
  B_{n} &= G_2B_{n-1},
  \label{eq:bnp1}\\
  G_1 &= G_2^{-1},
  \label{eq:Ginv}
\end{align}
where (\ref{eq:Ginv}) follows from the multiplication of (\ref{eq:G1_WTF}) and (\ref{eq:G2_WTF}). Equations (\ref{eq:anp1})-(\ref{eq:bnp1}) express the \emph{rheological property} of the platoon, that is, they define the form of how these two components propagate through the platoon. Equation (\ref{eq:Ginv}) expresses the \emph{principle of reciprocity}, that is, if $A(s)$ propagates with the help of $G_1(s)$ to higher indexes of vehicles, then $B(s)$ propagates with the help of $G_1(s)$ to lower indexes of vehicles. The function $G_1(s)$ is hereafter referred to as the wave transfer function.

It should be noted that if there is a boundary in the system, e.g., if the length of platoon is finite, where the rheology property for wave propagation changes abruptly, the principles must be supplemented by boundary conditions. We discuss this case in the following section.

\subsection{Verification of the wave transfer function}
We now outline an alternative way to derive the wave transfer function. Let the model of the vehicular platoon (\ref{eq:pos_decomp})-(\ref{eq:Ginv}) hold and now search for the transfer functions $G_1(s)$ and $G_2(s)$ that satisfy these four equations. Substituting (\ref{eq:pos_decomp}) into (\ref{eq:bidir_sys_laplace}) yields
\begin{align}
\alpha(A_n+B_n) = A_{n-1}+B_{n-1} + A_{n+1}+B_{n+1},
\end{align}
which, in view of (\ref{eq:anp1}) and (\ref{eq:bnp1}), is
\begin{align}
\alpha(s) = G_1(s) + G_2(s).
\end{align}
We can substitute either for $G_{1}(s)$ or $G_{2}(s)$ from (\ref{eq:Ginv}). Either possibility leads to the same quadratic equation ($m=1,2$),
\begin{align}
  G_m^2(s) - \alpha(s)G_m(s) +1=0,
  \label{eq:quadratic_equation}
\end{align}
with two linearly independent solutions,
\begin{align}
  G_{m}(s) = \frac{\alpha}{2} \mp \frac{1}{2}\sqrt{\alpha^2-4}.
  \label{eq:transferG_alpha}
\end{align}
Let $G_1(s)$ be chosen as the solution with the negative sign in front of the square root. Then (\ref{eq:Ginv}) only allows $G_2(s)$ to be the solution with the positive sign in front of the square root. Hence, $G_1(s)$ and $G_2(s)$ are identical to those derived in the previous section. The quadratic equation (\ref{eq:quadratic_equation}) can be employed as a starting model for the positioning of multi-link flexible mechanical systems \cite{OConnor2006}.

\subsection{Approximation of the wave transfer function}
It will be shown later in the paper that to be able to implement the wave-absorbing controller advertised at the beginning of the paper, we need to find the impulse response of the wave transfer function, i.e. the inverse Laplace transform of $G_1(s)$. Due to the presence of the square root in the function it is very challenging to find exact impulse response of $G_1(s)$. However, we can approximate the impulse response with a finite impulse response (FIR) filter. Therefore, we first approximate the wave transfer function in the Laplace domain, then transform this approximate form to the time domain and finally truncate and sample the approximate impulse response to obtain FIR filter coefficients.

The square root function in (\ref{eq:transferG_alpha}) can be approximated by various ways, e.g., Newton's method, the binomial theorem, or continued fraction expansion (\ref{eq:trans_continued_fraction}). We employ the last option since it guarantees the convergence of iterative approximations and is applicable to an arbitrary dynamics of the local system with a generalized parameter $\alpha(s)$ as in (\ref{eq:alpha_generalized}). The recursive formula (\ref{eq:trans_continued_fraction}) immediately provides the iterative approximation of $G_1(s)$,
\begin{equation}
  G_1^l(s) = \frac{1}{\alpha(s)-G_1^{l-1}(s)},
  \label{eq:G_recursive_approx}
\end{equation}
where $l = 1,2,\ldots $, and the initial value $G_1^0(s) = 1$. The approximate $G_1^l(s)$ can be transformed to the time domain by Matlab or Mathematica. Our experience with the inverse Laplace solvers for the Fractional Calculus \emph{invlap} \cite{dehoog1982}, \emph{weeks} \cite{Weeks1966} and \emph{nilt} \cite{brancik1999} in Matlab is that, while they were not capable of performing the inverse Laplace transform of (\ref{eq:transferG_alpha}) due to the square root function, they carried out the inverse Laplace transform of $G_{1}^l(s)$ without complications since (\ref{eq:G_recursive_approx}) is~a rational function.

\begin{figure}[htb]
 \centering
  \includegraphics[width=0.48\textwidth]{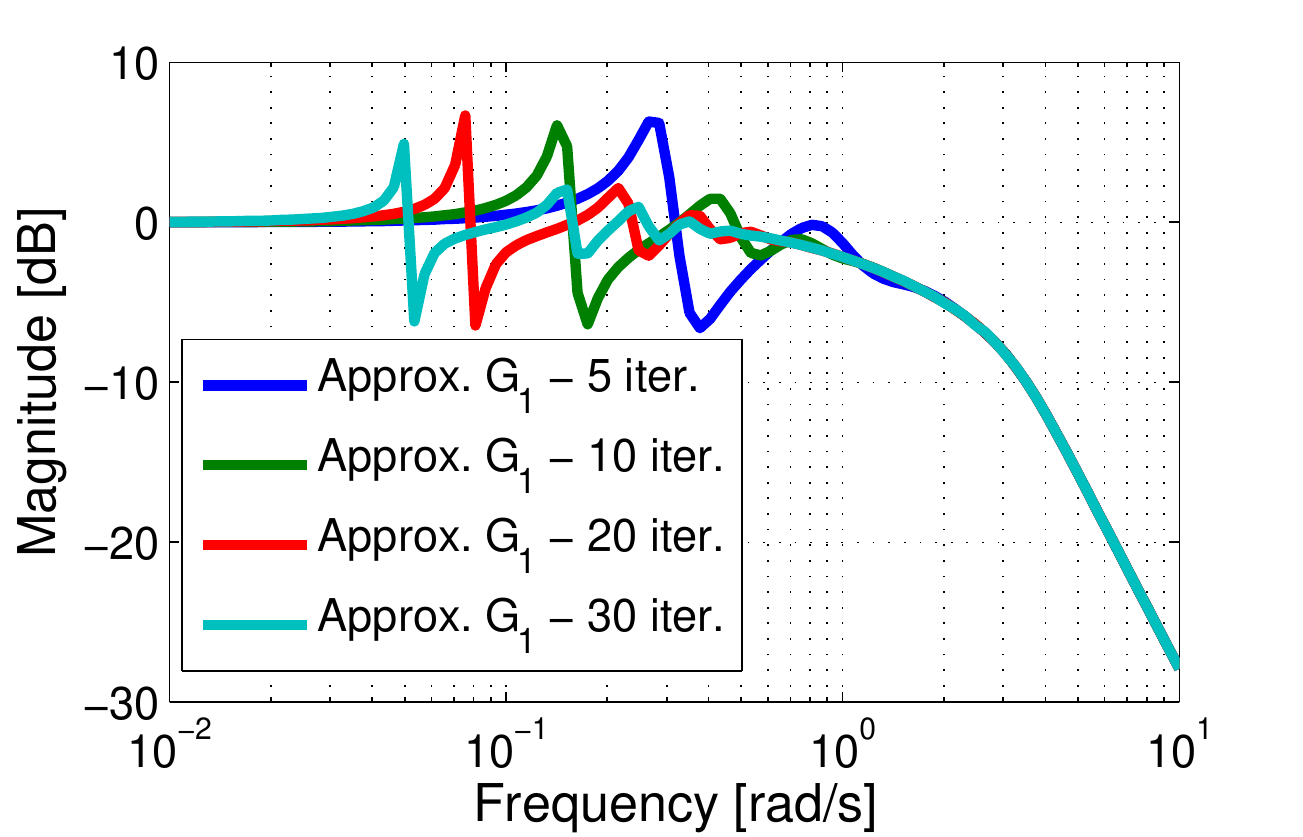}
  \caption{The Bode characteristics of $G_{1}(s)$ approximations after several iterations by (\ref{eq:G_recursive_approx}) for $k_{\text{p}}=k_{\text{i}}=\xi=4$.}
  \label{fig:bode_comparison_only_exact}
\end{figure}

\begin{figure}[htb]
 \centering
  \includegraphics[width=0.48\textwidth]{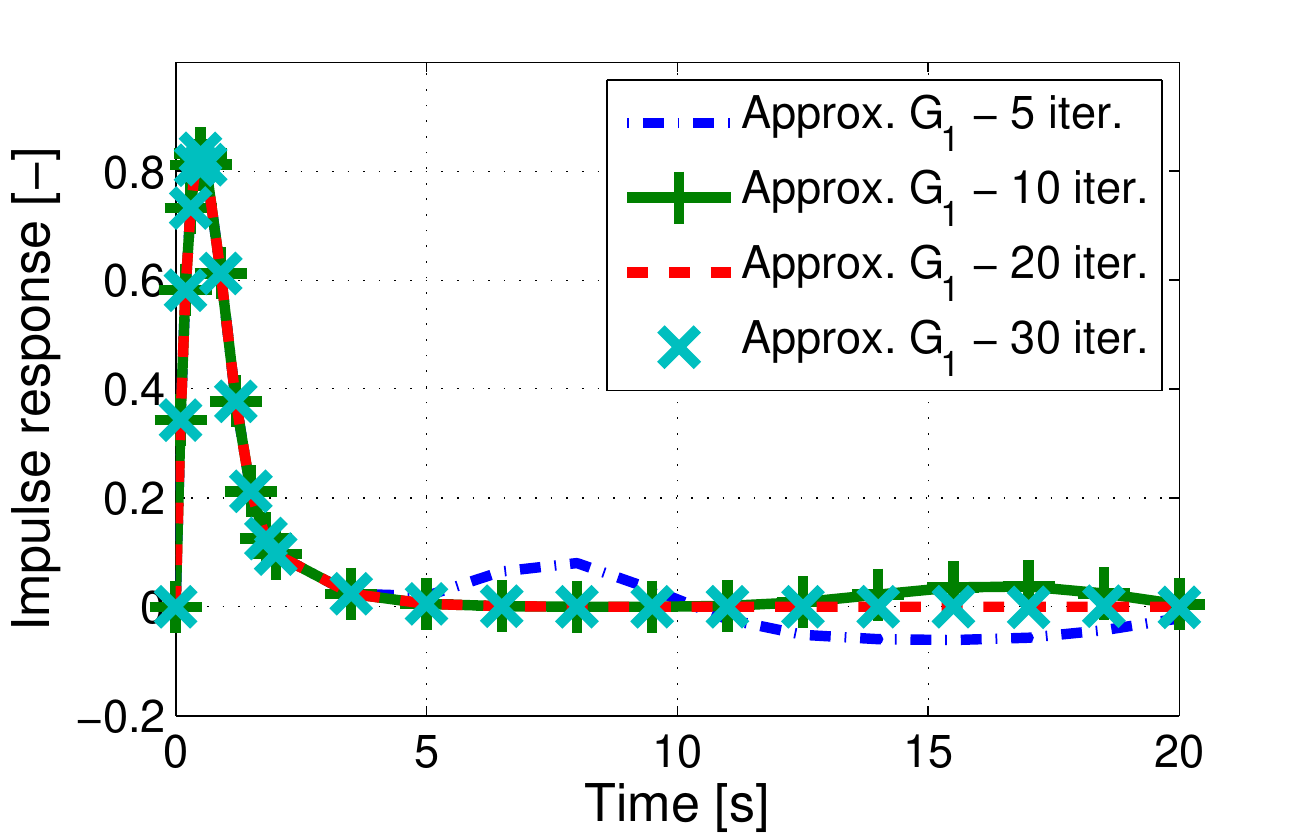}
  \caption{The impulse responses of $G_{1}(s)$ after several iterations by (\ref{eq:G_recursive_approx}) for $k_{\text{p}}=k_{\text{i}}=\xi=4$.}
  \label{fig:impulse_comparison_only_exact}
\end{figure}

The approximate $G_1^l(s)$ can interpreted as follows. Equation (\ref{eq:G_recursive_approx}) represents the transfer function from the position of the leader to the position of the first follower in a platoon of $l$ vehicles. Increasing the number of iterations (\ref{eq:G_recursive_approx}) means that the length of a platoon grows and the effect of the rear-end vehicle on $G_1(s)$ weakens. The approximation of $G_1(s)$ therefore successively improves. Figs. \ref{fig:bode_comparison_only_exact} and \ref{fig:impulse_comparison_only_exact} show the Bode characteristics $G_{1}^l(s)$ and the associated impulse responses for various number of iterations, respectively. Increasing the numbers of iterations makes the peak in the Bode characteristic sharper, more localized and moves it towards lower frequencies, eventually disappearing entirely. The basic characteristic of the impulse response is fitted after a few iterations while small differences occur at longer times. To obtain the FIR filter coefficients, we truncate the approximate impulse response at a few seconds and sample it with an appropriate frequency. In our numerical simulations it was sufficient to stop the iterative procedure after $20$ iterations, to truncate the impulse response at $15$ seconds and sample it at a frequency of $100\,\text{Hz}$.


\section{REFLECTION OF THE WAVE ON PLATOON ENDS}
\label{sec:reflections}

To be able to design a wave-absorbing controller for the platoon end, we first need to mathematically describe the wave reflection.

In the previous section an infinite platoon is considered, whereas here we assume a semi-infinite platoon having one end that is either externally controlled (forced end) or allowed to move freely (free end). When a wave propagates along a platoon and reaches its free end, it is reflected with the same polarity, i.e., the same sign of amplitude, but with the opposite polarity at the fixed/forced end. This phenomenon, known from basic wave physics \cite{french2003}, is discussed in the following in terms of the wave transfer function. The necessary mathematical derivations are given in Appendix A and B.

\subsection{The forced-end boundary}
\label{sec:forced_end_boundary}
We call the forced-end boundary such a vehicle that is externally controlled and is not to the other vehicles. However, the neighbouring vehicle is one-directionally linked with this forced boundary. The platoon leader therefore represents the forced-end boundary.
\begin{figure}[htb]
 \centering
 \includegraphics[width=0.49\textwidth]{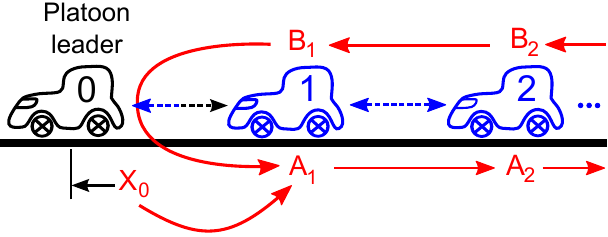}
  \caption{Scheme of wave reflection on the leader, i.e. reflection on the forced-end boundary, described by (\ref{eq:refl_fixed_end}).}
  \label{fig:fixed_end}
\end{figure}
The reflection on the forced-end boundary is sketched in Fig. \ref{fig:fixed_end}. Changing the position of the forced end, $X_0$, generates the outgoing wave as a first contribution to $A_1$. Moreover, the incoming wave ($B_1$) is reflected on the forced end and transformed to the outgoing wave as the second contribution to $A_1$. The force-end reflection is derived in \ref{sec:app_forced_end} and summarized by (\ref{eq:forced_end_appendix}),
\begin{equation}
  A_1 = G_1 X_{0}-G_1^2B_1.
  \label{eq:refl_fixed_end}
\end{equation}
This first shows that changing the position of the forced end is translated to $A_{1}$ through $G_{1}$. Second, since the DC gain of $G_1$ is equal to plus one (see Fig. \ref{fig:bode_comparison_only_exact}), the minus sign in front of $G_1^2$ causes the wave to be reflected with the opposite sign.

\subsection{The free-end boundary}
\label{sec:free_end_boundary}
A free-end boundary is a boundary where a vehicle is two-directionally linked with one neighbour only and, additionally, it is aware about steady state of the link. The rear-end vehicle described by (\ref{eq:last_sys_laplace}) represents the free-end boundary.

\begin{figure}[ht]
 \centering
  \includegraphics[width=0.49\textwidth]{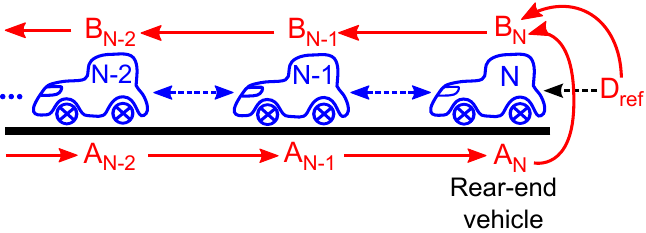}
  \caption{Scheme of wave reflection on the rear-end vehicle, i.e. reflection on the free-end boundary, described by (\ref{eq:refl_free_end}).}
  \label{fig:free_end}
\end{figure}

The reflection on the free-end boundary is outlined in Fig. \ref{fig:free_end}. The wave travelling from the free-end boundary ($B_N$) is composed of two parts, the incoming wave ($A_N$) which is reflected back through $G_1$ and the component due to adjusting the reference distance $D_{\text{ref}}$. The free-end reflection is derived in \ref{sec:app_free_end} and summarized by (\ref{eq:free_end_appendix}),
\begin{align}
B_N = G_1 A_N + \frac{G_1 -1}{\alpha-2} D_{\text{ref}}.
\label{eq:refl_free_end}
\end{align}
The reflection from the free-end boundary does not change the sign that is expressed by the plus sign in front of $G_1 A_N$. Moreover, the signal reflected from the free-end is delayed as a linear function of $G_{1}(s)$, while as a quadratic function when it is reflected from the forced-end boundary, as shown by (\ref{eq:refl_fixed_end}).

It should be noted that the verification of the above wave-based model was done in \cite{OConnor2007}. The transfer function
\begin{equation}
  \frac{X_N}{X_0} = G_1^N \frac{1+G_1}{1+G_1^{2N+1}},
\end{equation}
where $N$ is index of the last vehicle, was shown to be identical to the transfer function derived by the state space description. This result is valid not only for a double integrator with P controller, but for an arbitrary dynamics of the local system.

\section{WAVE-ABSORBING CONTROLLER}
\label{sec:wave-based_platoon_control}

The three main control requirements are: i) to travel the platoon at reference velocity $v_{\text{ref}}$, ii) to keep inter-vehicle distances $d_{\text{ref}}$, iii) to actively absorb the wave travelling towards the platoon's end.

This section introduces three possible configurations of the platoon with the wave-absorbing controller. First, we will describe the configuration where the wave-absorbing controller is implemented at the platoon leader.

\subsection{Front-sided wave-absorbing controller}
\label{sec:driving_ref_vel}

\subsubsection{Absorption of the wave}
\label{subsec:wave_absorption}

To absorb the incoming wave at the platoon front, the transfer function from $B_1$ to $A_1$ in (\ref{eq:refl_fixed_end}) has to be equal to zero. In other words, we are searching for $X_0$ to satisfy the equation $G_1 X_0/B_1 - G_1^2 = 0$. The only solution is
\begin{equation}
  X_0 = G_1B_1.
  \label{eq:leader_absorb}
\end{equation}
To be consistent with the model (\ref{eq:pos_decomp})-(\ref{eq:Ginv}), we denote $B_0 = G_1B_1$ and $A_0 = X_0-B_0$, then (\ref{eq:refl_fixed_end}) is expressed as $A_1 = G_1X_0-G_1B_0 = G_1 A_0$. Summarizing this yields the wave components of the leader
\begin{align}
  B_0 &= G_1X_1-G_1^2A_0, \label{eq:B0_leader}\\
  A_0 &= X_0-B_0.
\end{align}
This means that if one component of the position of the leader is equal to $B_0$, then the leader absorbs the incoming wave. We can imagine that if the leader is pushed/pulled by its followers, thus it manoeuvres like one of the in-platoon vehicles.

\subsubsection{Acceleration to the reference velocity}
\label{subsec:acceleration_platoon}

The previous algorithm actively absorbs the incoming wave to the platoon leader. To change the platoon's velocity and inter-vehicle distances are other tasks that need to be solved.

\begin{figure*}[htb]
 \centering
  \includegraphics[width=0.99\textwidth]{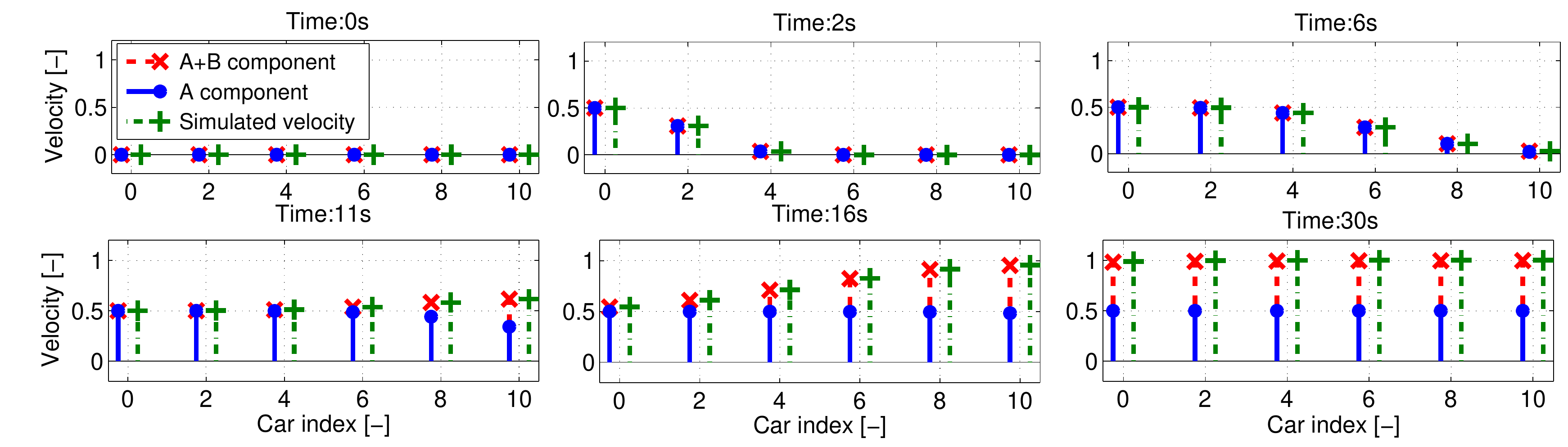}
  \caption{Simulation of the velocity wave propagating in the platoon with the Front-sided wave-absorbing controller at several time instances. At the beginning, $t=0\,{\text{s}}$, all platoon vehicles are standing still except for the leader which accelerates to a velocity $0.5\,\text{ms}^{-1}$. At intermediate times, the wave travels to the rear vehicle, where it is reflected and travels back to the leader to be completely absorbed. By propagating, it forces platoon vehicles to accelerate by another $0.5\,\text{ms}^{-1}$ to a velocity $1\,\text{ms}^{-1}$. At the final stage, $t=30\,{\text{s}}$, the leader is the last one reaching the velocity $1\,\text{ms}^{-1}$ and the whole platoon moves with $1\,\text{ms}^{-1}$. The red crosses represent the derivation of $A+B$ positional components computed by the wave transfer function approach, the green plus signs are the velocities simulated by the Matlab Simulink.}
  \label{fig:velocity_N10}
\end{figure*}

To accelerate the platoon, we need to add an external/reference input, $X_{\text{ref}}$, for the leader. This changes (\ref{eq:leader_absorb}) to $X_0 = B_0+X_{\text{ref}}$. The rear-end vehicle represents the free-end boundary, therefore, $B_0$ is expressed by the combination of (\ref{eq:anp1}), (\ref{eq:bnp1}), (\ref{eq:refl_fixed_end}) and (\ref{eq:B0_leader}) as $B_0 = G_1^{2N+1}X_{\text{ref}}$. This leads the transfer function from $X_{\text{ref}}$ to $X_0$ to be
\begin{equation}
  \frac{X_{\text{0}}}{X_{\text{ref}}} = 1+G_1^{2N+1}.
  \label{eq:leader_accel_trans}
\end{equation}
Fig. \ref{fig:bode_comparison_only_exact} showed that the DC gain of $G_1$ is equal to one, therefore, the DC gain of ($1+G_1^{2N+1}$) is equal to two. This means that to accelerate the platoon to reference velocity $v_{\text{ref}}$, the leader has to be commanded to accelerate to velocity $v_{\text{ref}}/2$ at the beginning of the manoeuver, as shown in Fig. \ref{fig:velocity_N10}.

Fig. \ref{fig:velocity_N10} additionally shows an independent validation of the wave transfer function approach. The derivation of the sum of $A+B$ velocity components (red crosses) of the wave travelling through the platoon are compared against the velocities simulated by the Matlab Simulink (green plus signs). We can see an agreement between the wave-transfer-function-derived and independently-simulated velocities.

\subsubsection{Changing of the inter-vehicle distances}
\label{subsec:changing_distances}

Increasing the inter-vehicle distances poses a more difficult task than merely accelerating the platoon. The reason is that the rear-end vehicle reacts to the change of reference distance $d_{\text{ref}}$ by acceleration/deceleration. This creates a velocity wave propagating towards the leader who absorbs it by changing its velocity. This means, however, that when all vehicles reach the desired inter-vehicle distance $d_{\text{ref}}$, the whole platoon travels with a new velocity different from the original. Only by an additional action of the leader, see the next paragraph, will the original velocity be reestablished.

Although the platoon has a finite number of vehicles, it behaves like a semi-infinite platoon because no wave reflects from the platoon leader, who is equipped with the wave absorber. Since (\ref{eq:refl_free_end}) holds for a semi-infinite platoon, it can be now used to determine the transfer function from $D_{\text{ref}}$ to velocity of the leader, $V_0(s)$, that is
\begin{equation}
  \frac{V_0}{D_{\text{ref}}} = G_1^N \frac{s(G_1-1)}{\alpha-2}.
  \label{eq:tf_Dref_V0}
\end{equation}
The DC gain of (\ref{eq:tf_Dref_V0}) reads as
\begin{equation}
   \kappa_{\text{f}} = \lim_{s\rightarrow 0} \left(G_1^N \frac{s(G_1-1)}{\alpha-2} \right).
\end{equation}

In the case where the reference distance is changed and the leader does not accelerate, the velocity of the platoon changes by $(\kappa_{\text{f}} d_{\text{ref}})$. This means that the platoon slows down or even moves backwards. To compensate for this undesirable velocity change, the leader is commanded to accelerate to the velocity
$(-\kappa_{\text{f}} d_{\text{ref}})/2$. The platoon will consequently travel with the original velocity, hence compensating for the acceleration/deceleration of the rear-end vehicle.

The DC gain of (\ref{eq:tf_Dref_V0}) for the PI controller case is equal to $(-\sqrt{k_{\text{i}}/\xi})$.

\subsubsection{Overall control of the leader}
\begin{figure*}[htb]
 \centering
  \includegraphics[width=0.8\textwidth]{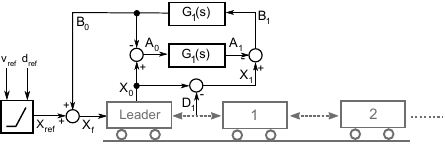}
  \caption{Scheme of the Front-sided wave-absorbing vehicular platoon controller.}
  \label{fig:platoon_actuators_one}
\end{figure*}

Let us now assume that the leader has a positional controller with input $X_{\text{f}}$. Summarizing preceding subsections yields the resulting control law of the leader,
\begin{align}
X_{\text{f}}(s) &= X_{\text{ref}}(s) + B_0(s),
\label{eq:external_input_leader}
\end{align}
From the above discussion, $X_{\text{ref}}(s)$ must be represented by a ramp signal with slope $w_{0}$,
\begin{equation}
w_{0} = \frac{1}{2}\left(v_{\text{ref}}-\kappa_{\text{f}} d_{\text{ref}}\right),
\label{eq:ramp_leader}
\end{equation}
to ensure that the platoon travels with a reference velocity $v_{\text{ref}}$ and inter-vehicle distances $d_{\text{ref}}$. In case of the PI controller, $w_0 = \left(v_{\text{ref}} +\sqrt{k_{\text{i}}/\xi} d_{\text{ref}}\right)/2$. The Front-sided wave-absorbing controller is summarized in Fig. \ref{fig:platoon_actuators_one}.

\subsection{Rear-sided wave-absorbing controller}
\label{sec:rear_sided_WBC}

Instead of placing the wave-absorbing controller at the platoon's front, it can be placed at the platoon's rear. In this case, the platoon has one leader in the front and one wave-absorbing controller at the rear. However, the absence of the predecessor follower in the platoon has an important consequence. Any velocity change of the leader, $V_{0}(s)$, causes a change in the distance to the first follower, $D_1(s)$, as shown in (\ref{eq:forced_end_dist_vel}). Consequently, all other distances between vehicles are changed. This negative effect is to be compensated by an acceleration/deceleration of the rear-end vehicle. We denote $\kappa_{\text{r}}$ to be the DC gain of the transfer function from $V_0(s)$ to $D_1(s)$.

Having specified the DC gain, a certain reference signal needs to be sent to the platoon end to set up a desired inter-vehicle distance $d_{\text{ref}}$. The input to the positional controller of the rear-end vehicle, $X_{\text{r}}(s)$, is expressed, analogous to (\ref{eq:external_input_leader}), as
\begin{align}
X_{\text{r}}(s) &= X_{\text{ref,rear}}(s) + G_1(s)A_{N-1}(s),
\label{eq:external_input_last}
\end{align}
where $X_{\text{ref,rear}}(s)$ is a reference ramp signal with slope $w_{r}$,
\begin{equation}
w_{r} = \frac{1}{2}\left(v_{\text{ref}}-\kappa_{\text{r}} d_{\text{ref}}\right).
\label{eq:ramp_last}
\end{equation}
In other words, the platoon leader drives the platoon to travel with velocity $v_{\text{ref}}$, while the rear-end vehicle makes the platoon travel with inter-vehicle distances $d_{\text{ref}}$. For the PI controller case $\kappa_{\text{r}} = k_{\text{i}}/\xi$.

\subsection{Two-sided wave-absorbing controller}
\label{sec:two_sided_WBC}

The Front-sided and Rear-sided wave-absorbing controllers can be combined by implementing wave absorbers to both the platoon leader and the rear-end vehicle. In this case, no wave is reflected back from neither of platoon ends.

The input to the positional controller of the leader is given by (\ref{eq:external_input_leader}) with the ramp signal (\ref{eq:ramp_leader}), while the input to the positional controller of the rear-end vehicle is (\ref{eq:external_input_last}) with the ramp signal (\ref{eq:ramp_last}). In this way, each platoon end generates a velocity wave propagating towards the opposite end. Likewise, as for the Front-sided and Rear-sided wave-absorbing controllers (Section \ref{sec:driving_ref_vel} and \ref{sec:rear_sided_WBC}), the amplitudes of the two waves are summed up to $v_{\text{ref}}$, meaning that the platoon travels with velocity $v_{\text{ref}}$ and inter-vehicle distances $d_{\text{ref}}$.

\subsection{Asymptotic and string stability}
Using the same technique as in \cite{Herman2013}, it can be shown that a platoon with the symmetric bidirectional controller is asymptotically stable. Since $G_1(s)$ can be represented by such a platoon, it is asymptotically stable as well. The truncated approximate of $g_1(t)$ is BIBO (bounded-input bounded-output) stable, which is a well known fact about FIR filters. Therefore, a platoon with the wave-absorbing controller on one or both platoon ends remains asymptotically stable.

We follow the $L_2$ string stability definition from \cite{Eyre1998a} that can be formulated as:\emph{ The system is called $L_2$ string stable if there is an upper bound on the $L_2$-induced system norm of $T_{0,n}$ that does not depend on the number of vehicles, where $T_{0,n}$ is the transfer function from position of the leader to the position of the vehicle indexed $n$.}

In the case of the platoon with the Front-sided wave-absorbing controller, the position of the $n$th vehicle is described as
\begin{equation}
  X_{n} = (G_1^{n}+G_1^{2N+1-n}) X_{0}.
\end{equation}
Due to the triangle inequality and the fact that $||G_1||_{\infty} \leq 1$, which is shown in \ref{app:string_stability}, we obtain
\begin{equation}
  ||G_1^{n}+G_1^{2N+1-n}||_{\infty} \leq ||G_1^{n}||_{\infty}+||G_1^{2N+1-n}||_{\infty} \leq 2.
\end{equation}
This means that the magnitude of the maximum peak in the frequency response of the transfer function from the position of the leader to the position of the $n$th vehicle is smaller or equal to $2$. Since the $L_2$-induced norm and $H_{\infty}$ coincide, we can state that the platoon with the Front-sided wave-absorbing controller is $L_2$ string stable.

The position of the $n$th vehicle with an absorber placed at the rear-end vehicle is
\begin{equation}
  X_{n} = G_1^{n}X_0 + (G_1^{N-n}-G_1^{N+n})X_{N}.
\end{equation}
We apply the same idea and state that $H_{\infty}$ norm of both $G_1^{n}$ and $(G_1^{N-n}-G_1^{N+n})$ are bounded regardless of the number of vehicles. Therefore, the platoon with the Rear-sided wave-absorbing control is $L_2$ string stable.

The position of the $n$th vehicle in a platoon with absorbers on both ends is expressed as
\begin{equation}
  X_n = G_1^n X_{0} +G_1^{N-n}X_{N},
\end{equation}
which immediately shows that the platoon with the Two-sided wave-absorbing controller is $L_2$ string stable as well.

\section{NUMERICAL SIMULATIONS}
We consider the linear friction of our system to be $\xi = 4$ and search for the parameters of the PI controller such that oscillations of the impulse response of $G_1(s)$ are minimized. The parameters $k_{\text{p}} = k_{\text{i}} = 4$ satisfy this requirement. All numerical simulations are run for a platoon of $50$ vehicles to demonstrate that the wave-absorbing controllers are capable of controlling large platoons.

\begin{figure*}[!htbp]
 \centering
   \includegraphics[width=0.99\textwidth]{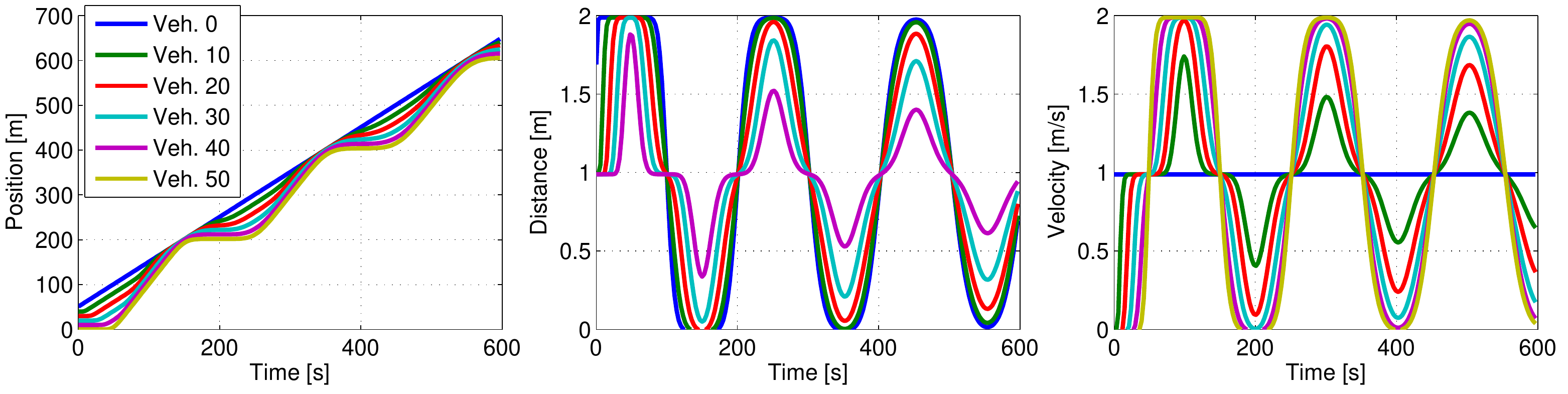}
  \caption{Simulation of the platoon without the wave-absorbing controller when the leader accelerates to velocity $v_{\text{ref}} = 1\,\text{ms}^{-1}$. The reference distance is kept fixed, $d_{\text{ref}} = 1\,\text{m}$, for the whole time.}
  \label{fig:accel_of_platoon_noWBC}
\end{figure*}

To demonstrate the advantages of the wave-absorbing controllers, we will compare their performance against a pure bidirectional control without any wave-absorbing controller. This means that the leader travels with a constant velocity $v_{\text{ref}}$ for the whole time of the simulation. Fig. \ref{fig:accel_of_platoon_noWBC} shows outcomes of numerical simulation when the leader without wave-absorbing controller increases its velocity. We can see significant limitations of the bidirectional control. The oscillatory behaviour in the movement of the platoon is caused by numerous wave reflections from both platoon ends. Eventually, the platoon settles at a desired velocity after many velocity oscillations. These oscillations not only significantly prolong the settling time, but they could lead to accidents within the platoon.

\begin{figure*}[!htbp]
 \centering
  \includegraphics[width=0.99\textwidth]{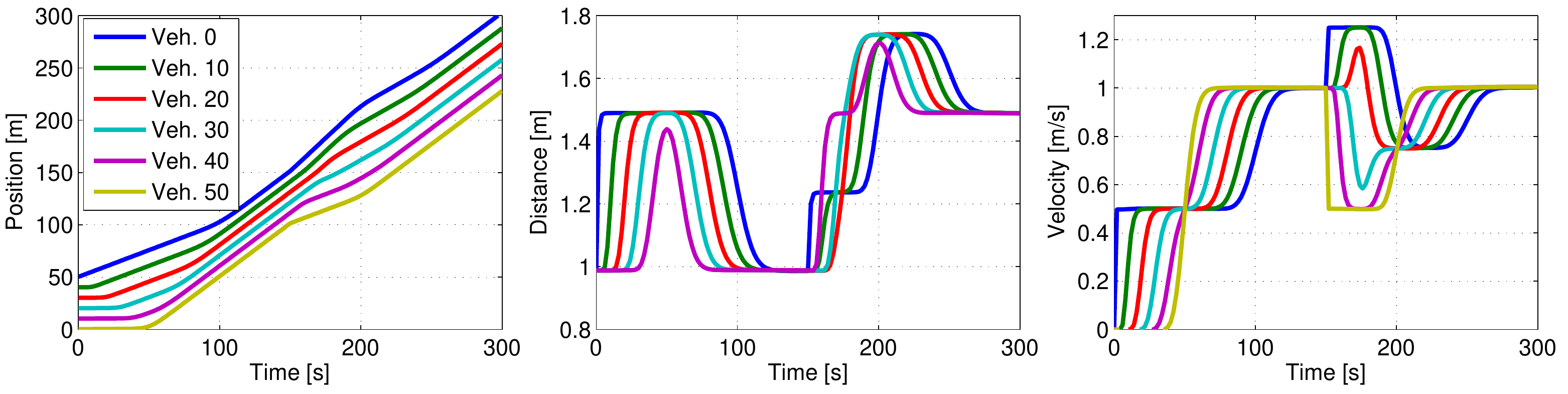}
  \caption{Simulation of two platoon manoeuvrers with the Front-sided wave-absorbing controller. At the beginning, the vehicles are standing still separated by one meter. For the first manoeuver, the platoon is commanded to accelerate to $v_{\text{ref}}=1\,\text{ms}^{-1}$ with $d_{\text{ref}} = 1\,\text{m}$ starting at time $t=0\,\text{s}$.  At time $t = 150\,\text{s}$, the platoon is commanded to perform the second manoeuver such that the reference distance is increased to $d_{\text{ref}} = 1.5\,\text{m}$ without changing the reference velocity.}
  \label{fig:accel_of_platoon_WBC}
\end{figure*}

The performance of the Front-sided wave-absorbing controller during two platoon manoeuvrers is shown in Fig. \ref{fig:accel_of_platoon_WBC}. In the first $150\,\text{s}$ manoeuver, the platoon accelerates (not necessarily from zero velocity) to reach a desired velocity. In comparison with the pure bidirectional control, see Fig. \ref{fig:accel_of_platoon_noWBC}, the settling time is now significantly shorter. Moreover, under some circumstances, it can be guaranteed that vehicles do not crash into each other during the platoon acceleration. In fact, the distances between vehicles are increased at the beginning of the acceleration as suggested by (\ref{eq:forced_end_dist_vel}) and shown in the middle panel of the Fig. \ref{fig:accel_of_platoon_WBC}. However, the distances may undershoot the initial inter-vehicles distances in the second part of the acceleration manoeuver. If the impulse response of the wave transfer function is tuned such that it does not undershoot the zero value, then the distances between vehicles can not become less than the initial inter-vehicle distances. In the opposite case (not shown here), where the platoon travels with a constant velocity and starts to decelerate, the distances between vehicles are temporarily decreased and a collision may occur.

At time $t = 150\,\text{s}$ in Fig. \ref{fig:accel_of_platoon_WBC}, the platoon is commanded to perform the second manoeuver such that the reference distance is increased, but the reference velocity is kept unchanged. The rear-end vehicle reacts to this command at the same time as the leader since it is controlled by the reference distance that is now changing. However, the end vehicles differ in action; the leader accelerates, while the rear-end vehicle decelerates. This behaviour creates an undesirable overshoot in distances.

\begin{figure*}[!htbp]
 \centering
  \includegraphics[width=0.99\textwidth]{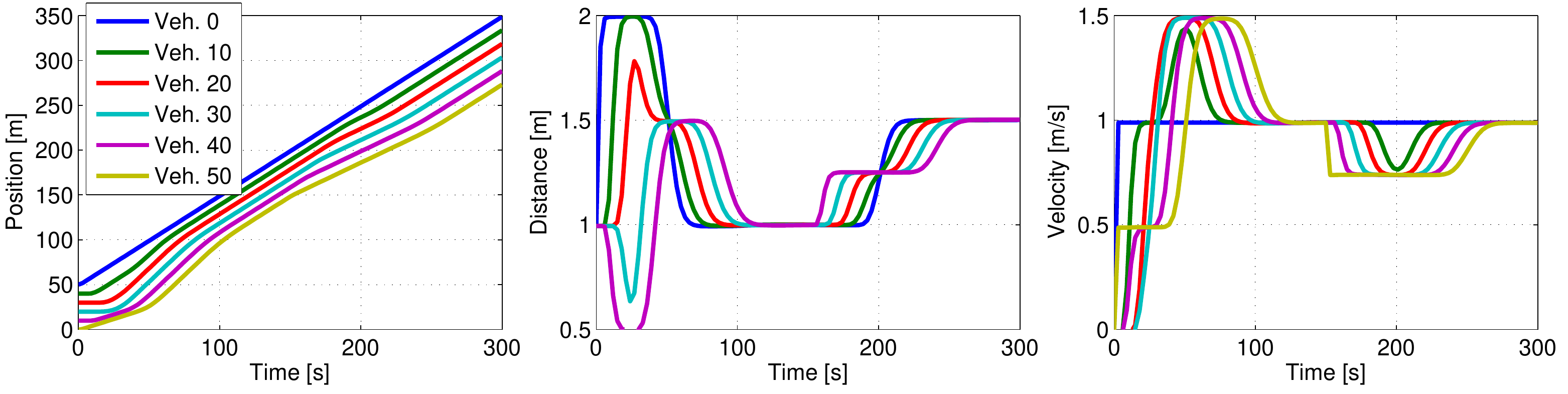}
    \caption{As in Fig. \ref{fig:accel_of_platoon_WBC} but with the Rear-sided wave-absorbing controller.}
  \label{fig:rear_sided_WBC}
\end{figure*}

A numerical simulation of the two manoeuvrers for the platoon controlled by the Rear-sided wave-absorbing controller is shown in Fig. \ref{fig:rear_sided_WBC}. During the acceleration manoeuver the inter-vehicle distances between vehicles closer to the rear end are temporarily decreased while those for vehicles near the leader are temporarily increased. During the changing-distance manoeuver, on the other hand, no overshoot in distances occurs.

\begin{figure*}[!htbp]
 \centering
  \includegraphics[width=0.99\textwidth]{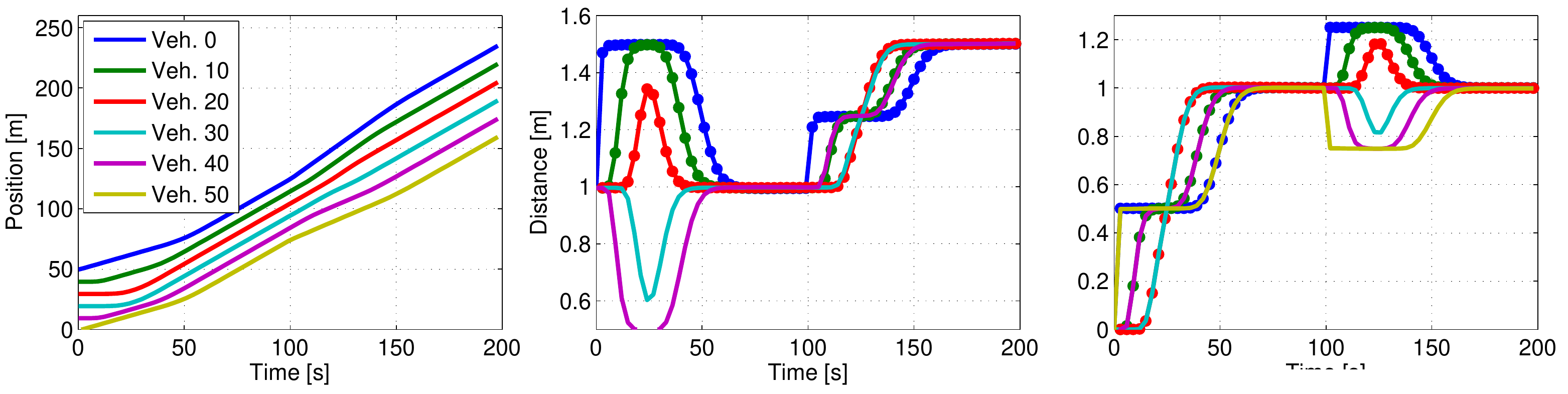}
    \caption{As in Fig. \ref{fig:accel_of_platoon_WBC} but with the Two-sided wave-absorbing controller. The second command to increase $d_{\text{ref}}$ comes at $t = 100\,{\text{s}}$.}
  \label{fig:two_sided_WBC}
\end{figure*}

In Fig. \ref{fig:two_sided_WBC}, the acceleration and changing-distance manoeuvrers carried out for the one-sided wave-absorbing controllers are now performed for the two-sided wave-absorbing controller. Since both platoon ends are fully controlled, the settling time is only half of that for the one-sided wave-absorbing controllers. The middle panel in Fig. \ref{fig:two_sided_WBC} shows that there is no overshoot in distances during the second manoeuver. On the other hand, there is no guarantee that the vehicles will not collide during the acceleration manoeuver.

\subsection{Evaluation of the performance}
We now evaluate the performance of the acceleration manoeuver described in the previous section with the help of the mean squared error (MSE) criterion,
\begin{equation}
  \text{MSE} = \frac{1}{N+1} \sum_{n=0}^{N} \frac{1}{T}\sum_{t=0}^{T} (v_{\text{ref}}(t) -v_n(t))^2,
  \label{eq:MSE}
\end{equation}
where $T$ is the simulation time (in our case $T=500\,\text{s}$), $v_{\text{ref}}(t)$ is the reference velocity of the platoon at time $t$ and $v_n(t)$ is the actual velocity of the $n$th vehicle at time $t$.

\begin{figure}[ht]
 \centering
  \includegraphics[width=0.49\textwidth]{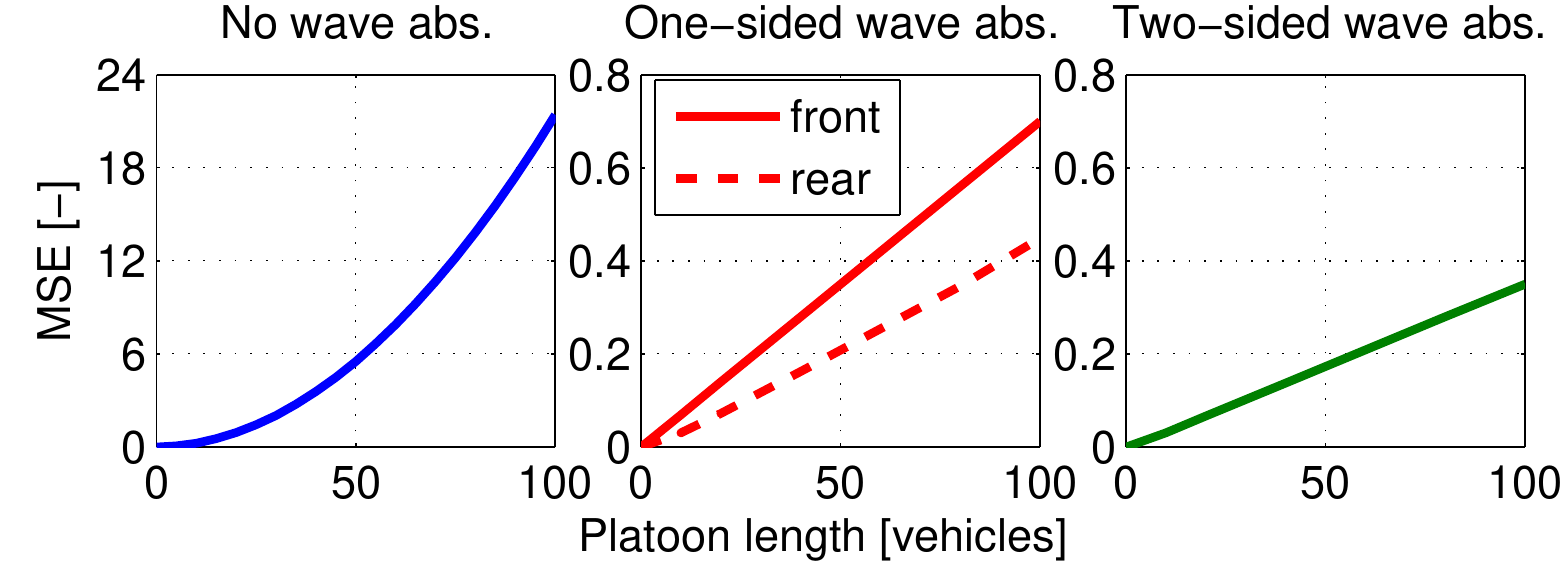}
  \caption{MSE performance evaluation of the acceleration manoeuver from Figs. \ref{fig:accel_of_platoon_noWBC}, \ref{fig:accel_of_platoon_WBC} and \ref{fig:two_sided_WBC}. All the four controllers are evaluated; pure bidirectional without wave absorber (left panel), Front-sided wave-absorbing controller (solid line in the middle panel), Rear-sided wave-absorbing controller (dashed line in the middle panel) and Two-sided wave-absorbing controller (right panel) for various platoon lengths according to (\ref{eq:MSE}).}
  \label{fig:stats_accel}
\end{figure}

The comparison in performance of the four controllers for various platoon lengths is depicted in Fig. \ref{fig:stats_accel}. We can see that the MSE increases linearly for all wave-absorbing controllers, but quadratically for the pure bidirectional control without wave absorber. Moreover, a linear increase in MSE for the Two-sided controller is only about half of that for the Front-sided controller. The linear increase of MSE for the Rear-sided controller lies between these two cases. Evidently, the wave-absorbing controller qualitatively improves the performance of the bidirectional control.

The settling time of the acceleration manoeuver arising from the four types of controllers are compared in Table \ref{table_settling_time}. We can see that the settling time increases quadratically with the platoon length for a platoon without wave-absorbing controller, but approximately linearly for a platoon with wave-absorbing controllers.
\begin{table}[!h]
\renewcommand{\arraystretch}{1.1}
\caption{The time required the platoons of various lengths to accelerate and stay within a range of $5\%$ of $v_{\text{ref}}$.}
\label{table_settling_time}
\centering
\small
\begin{tabular}{|c|c|c|c|c|}
\hline
& No wave &  Front-sided & Rear-sided & Two-sided\\
& abs.& wave abs. & wave abs. & wave abs. \\
\hline
 5 veh. & $70\,\text{s}$ & $12\,\text{s}$ & $11\,\text{s}$ & $7.5\,\text{s}$\\
\hline
 10 veh. & $322\,\text{s}$ & $24\,\text{s}$ & $23\,\text{s}$ & $14\,\text{s}$\\
\hline
20 veh. & $1365\,\text{s}$ & $46\,\text{s}$ & $45\,\text{s}$ & $26\,\text{s}$\\
\hline
 40 veh. & $5460\,\text{s}$ & $90\,\text{s}$ &$88\,\text{s}$& $49\,\text{s}$\\
\hline
\end{tabular}
\end{table}
\normalsize

\subsection{Effect of noise in the platoon}

This subsection examines the performance of the four controllers when noise is present in the system. The reference commands for a platoon of $20$ vehicles are $v_{\text{ref}}= 0\,\text{ms}^{-1}$ and $d_{\text{ref}} = 0\,\text{m}$, that is, the platoon is commanded not to move. Normally distributed noise is simulated for $2000$ seconds and added to distance measurements of each vehicle, except for the leader. Different realizations of a normally distributed noise with the mean value $\mu = 0$ and variance $\sigma^2 = 1$ are applied to each vehicle.

Table \ref{table_noise} assesses quantitatively the effect of noise on the performance of the four controllers. The mean squared error of positions, $\text{MSE}_{\text{pos}}$, and the arithmetic mean of positions, $\text{Mean}_{\text{pos}}$, show that the platoon without any absorber and with the Rear-sided wave-absorbing controller perform significantly better than with the other two controllers. This is due to the fact that, at least, one of the platoon ends is anchored at position $0$, meaning that the platoon does not drift away from position $0$, which is not the case for the Front-sided and Two-sided wave-absorbing controllers. Despite the disturbances by noise, all wave-absorbing controllers are better at maintaining in the coherence of the platoon than the pure bidirectional controller, as indicated by the mean squared error of inter-vehicle distances, $\text{MSE}_{\text{dist}}$, and the maximum distance between the leader and the rear end, $\text{MAX}_{\text{dist}}$.

\begin{table}[!h]
\renewcommand{\arraystretch}{1.1}
\caption{Performance of the four controllers when considering normally distributed noise affecting distance measurement of vehicles. Four criterions used for evaluation are introduced in the text.}
\label{table_noise}
\centering
\small
\begin{tabular}{|c|c|c|c|c|}
\hline
& No wave & Front-sided &  Rear-sided & Two-sided\\
& abs. & wave abs. & wave abs. & wave abs. \\
\hline
 $\text{MSE}_{\text{pos}}$ & $2.7\times 10^7$ & $8.4\times 10^7$ & $7.1\times 10^5$ & $1.3\times 10^8$\\
\hline
 $\text{Mean}_{\text{pos}}$ & $2\times 10^{-3}$ & $-3.9$ & $3.2\times 10^{-3}$ & $-3.1$\\
\hline
$\text{MSE}_{\text{dist}}$ & $1.9\times 10^5$ & $2.4\times 10^4$ & $2.5\times 10^4$ & $1.8\times 10^4$\\
\hline
$\text{MAX}_{\text{dist}}$ & $5.75$ & $1.37$ & $1.15$ & $0.64$\\
\hline
\end{tabular}
\end{table}
\normalsize

\section{CONCLUSIONS}
This paper introduces novel concepts for the control of a vehicular platoon, which significantly improve the popular bidirectional control. The main idea is to control the front end or both ends of a platoon to actively damp the waves of positional changes arriving from the opposite platoon end. The absorbing-end vehicle is assumed to i) measure the distance to its neighbour, ii) know its own position and iii) represent the dynamics of a vehicle in terms of the wave transfer function.

The new schemes allow us to control the platoon velocity and the inter-vehicle distances without long-lasting transient and oscillatory behaviour. The velocity errors during the platoon manoeuvres with the traditional bidirectional control grows quadratically with number of vehicles in the platoon, while errors grows only linearly for the bidirectional control enhanced with the wave-absorbing controller. Moreover, the platoon with the wave-absorbing controller is string stable. 

Additionally, the wave-absorbing controller preserves advantages of the bidirectional control such as: i) The lack of a need for vehicle-to-vehicle communication, ii) none of the vehicles needs to know the number of vehicles in the platoon, iii) an in-platoon vehicle does not need to know its relative position in the platoon, and iv) an in-platoon vehicle does not need to know the reference velocity and the reference distance for the platoon.

However, a considerable mathematical difficulty in the wave-absorbing control lies in finding the impulse response of the wave transfer function. In this paper, we proposed the iterative approach of constructing an approximation of the wave transfer function that is based on a continued fraction representation. Even for a small number of iterative steps, when the wave transfer function is rather roughly approximated, the wave-absorbing control still performs efficiently to damper oscillations in the platoon's characteristics (i.e. velocity, inter-vehicle distances).

It should be noted that the absorbing-end vehicle is assumed to be equipped with the positional controller since the differences in positions between vehicles are controlled. Alternatively, when the absorbing-end vehicle is equipped with a velocity controller, the commanded position of the vehicle derived using (\ref{eq:external_input_leader}) or (\ref{eq:external_input_last}) can be numerically differentiated to obtain the velocity commanded to the absorbing-end vehicle.

Undesirable overshoots in the velocities or inter-vehicle distances of the wave-absorbing control can be eliminated by introducing time delays in the reference signal applied to one of the platoon ends. An appropriate value of this time delay is dependent upon the platoon length and thus requires the extension of the wave-absorbing control. This topic warrants further investigation.

This paper extends \cite{Martinec2014} submitted on 15.~October 2013 in the following way: i) It presents mathematical derivation of the approximating formula for the wave transfer function and derivation of the transfer functions describing wave reflection on platoon ends, ii) it generalizes the result from double integrator model with linear friction and PI controller for an arbitrary local system dynamics, iii) it introduces two additional modifications of the wave-absorbing controller for the vehicular platoon, iv) it analyses asymptotic and string stability of a platoon with the wave-absorbing controller and v) it more thoroughly evaluates performance of the wave-absorbing controller.

\section{ACKNOWLEDGEMENTS}
This work was supported by Grant Agency of the Czech Republic within the project GACR P103-12-1794.

The authors thank Kevin Fleming for his comments on the manuscript.

\appendix
\section{Reflection on a forced-end boundary}
\label{sec:app_forced_end}
In this appendix, we derive the formula describing the reflection of a wave on a forced-end boundary that is defined in Section \ref{sec:forced_end_boundary}.

We first combine (\ref{eq:pos_decomp})-(\ref{eq:bnp1}) to obtain
\begin{align}
  X_{n+1} &= G_{1}A_n + G_2B_n, \label{eq:X_np1}\\
  X_{n-1} &= G_{2}A_n + G_1B_n. \label{eq:X_nm1}
\end{align}
Equation (\ref{eq:bidir_sys_laplace}) specified for the first vehicle behind the platoon leader is therefore
\begin{equation}
  \alpha X_1 = X_{0} + X_2.
\end{equation}
Substituting (\ref{eq:pos_decomp}) for $X_1$  and (\ref{eq:X_np1}) for $X_2$ yields
\begin{equation}
  \alpha (A_1 + B_1) = X_{0} + G_1A_1 + G_2B_1,
\end{equation}
which can be reformulated as
\begin{equation}
  A_1 = \frac{1}{\alpha - G_1}X_{0} +\frac{G_2 - \alpha}{\alpha-G1} B_1.
  \label{eq:A1_fixed_proof1}
\end{equation}
The term in front of $B_1$ can be arranged as
\begin{align}
  \frac{G_2 - \alpha}{\alpha-G1} = \frac{-\frac{\alpha}{2}+\frac{1}{2}\sqrt{\alpha^2-4}}{\frac{\alpha}{2} + \frac{1}{2}\sqrt{\alpha^2-4}} = -\frac{G_1}{G_2} = -G_1^2,
\end{align}
where the principle of reciprocity from (\ref{eq:Ginv}) has been applied. Similarly, the term in front of $X_{0}$ is expressed as
\begin{align}
  \frac{1}{\alpha-G1} = \frac{1}{\frac{\alpha}{2} + \frac{1}{2}\sqrt{\alpha^2-4}} = \frac{1}{G_2} = G_1.
\end{align}
Finally, we have
\begin{equation}
  A_1 = G_1 X_{0} - G_1^2 B_1.
  \label{eq:forced_end_appendix}
\end{equation}

The wave-based platoon control in Section \ref{sec:two_sided_WBC} requires one to specify the way how the velocity of the leader $V_0(s)$ influences the distance to the first follower $D_1(s)$, $D_{1}(s) = X_{0}(s) - X_1(s)$. Assuming a semi-infinite platoon, equation (\ref{eq:G1_continued_fraction}) gives $X_1(s) = G_1(s) X_{0}(s)$. Hence,
\begin{align}
  D_{1}(s) = X_{0}(s) - G_1(s) X_{0}(s) = \frac{1}{s}(1-G_1(s)) V_0(s),
\end{align}
In other words, the transfer function from velocity $V_{0}(s)$ to distance $D_{1}(s)$ is
\begin{equation}
  \frac{D_1(s)}{V_0(s)} = \frac{1}{s}(1-G_1(s)).
  \label{eq:forced_end_dist_vel}
\end{equation}

\section{Reflection on a free-end boundary}
\label{sec:app_free_end}
In this appendix, we derive the formula describing the reflection of a wave on a free-end boundary that is defined in Section \ref{sec:free_end_boundary}.

Substituting (\ref{eq:pos_decomp}) and (\ref{eq:X_nm1}) into (\ref{eq:last_sys_laplace}) yields
\begin{equation}
  (A_N+B_N) (\alpha -1) = G_2A_N+G_1B_N -D_{\text{ref}},
\end{equation}
which, after rearranging, gives
\begin{equation}
  B_N = \frac{G_2-\alpha+1}{\alpha-1-G_1}A_N - \frac{1}{\alpha-1-G_1}D_{\text{ref}},
  \label{eq:BN_free_proof1}
\end{equation}
where
\begin{align}
  &\frac{G_2-\alpha+1}{\alpha-1-G_1} = \frac{1-\frac{\alpha}{2}+\frac{1}{2}\sqrt{\alpha^2-4}} {-1+\frac{\alpha}{2}+\frac{1}{2}\sqrt{\alpha^2-4}} = \nonumber \\ &\frac{\alpha-\frac{\alpha^2}{2}- \sqrt{\alpha^2-4} +\frac{\alpha}{2}\sqrt{\alpha^2-4}}{2-\alpha} = \frac{2-\alpha}{2-\alpha} G_1 = G_1.
\end{align}
Similarly,
\begin{align}
  &\frac{1}{\alpha-1-G_1} = G_1 \frac{1}{G_2-\alpha+1} = \nonumber \\ &\left(\frac{\alpha}{2}-\frac{1}{2}\sqrt{\alpha^2-4} \right)\frac{1-\frac{\alpha}{2}+\frac{1}{2}\sqrt{\alpha^2-4}}{\left(1-\frac{\alpha}{2} \right)^{2} -\frac{1}{4}(\alpha^2-4)} = \frac{G_1 -1}{2-\alpha}.
\end{align}
Hence, (\ref{eq:BN_free_proof1}) is
\begin{equation}
  B_N = G_1 A_N + \frac{G_1 -1}{\alpha-2} D_{\text{ref}}.
  \label{eq:free_end_appendix}
\end{equation}

\section{PROOF OF $||G_1(s)||_{\infty} \leq 1$}
\label{app:string_stability}

We will show that $||G_2(s)||_{\infty} \geq 1$. Then (\ref{eq:Ginv}) implies that $||G_1(s)||_{\infty} \leq 1$. To inspect the amplification and phase shift on frequency $\omega$, we substitute $\jmath\omega$ for $s$ in the definition of $G_2(s)$ in (\ref{eq:G2_WTF}), where $\jmath$ is the imaginary unit, and obtain the complex number $z_2$ in the polar form,
\begin{equation}
  z_2 = r_2 \exp(\jmath \varphi_2).
\end{equation}
Similarly as in (\ref{eq:G2_WTF}), we can separate $z_2$ into two parts,
\begin{align}
  z_2 = \frac{1}{2}z + \frac{1}{2}\sqrt{z_s},
\end{align}
where
\begin{align}
z &= \alpha(\jmath\omega) = r \exp(\jmath \varphi), \nonumber\\
z_s &= z^2-4 = r_s \exp(\jmath \varphi_s).
\label{eq:app_zs}
\end{align}
The magnitude $r_s$ is given by
\begin{align}
  r_s &=  (r^2\cos(2\varphi)-4)^2 + (r^2\sin(2\varphi))^2 \nonumber \\
  &= r^4+8r^2+16-16r^2\cos^2\varphi.
\end{align}
with magnitude $r_2$ then expressed as
\begin{align}
  r_2 &=
   \left[ \frac{1}{4} \left(r^2 +r_s +2r\sqrt{r_s} \left(\cos\frac{\varphi_s}{2}\cos\varphi+\sin \frac{\varphi_s}{2}\sin\varphi\right) \right) \right]^{\frac{1}{2}} \nonumber\\
  &= \left[\frac{1}{4} \left(r^2 +r_s +2r\sqrt{r_s} \cos\left(\varphi-\frac{\varphi_s}{2} \right)\right) \right]^{\frac{1}{2}}.\label{eq:app_r2}
\end{align}
The minimum of $r_s$ over all possible phases is for $\varphi = k \pi$, $k\in \mathbb{Z}$, and is equal to
\begin{equation}
  \min(r_{\text{s}}) = \sqrt{r^4-8r^2+16} = |r^2-4| =
  \begin{cases}
    4-r^2 & \text{if } 0 \leq r \leq 2 \\
    r^2-4 & \text{if } r > 2
  \end{cases}
\end{equation}
Therefore,
\begin{equation}
  \frac{1}{4}(r^2+r_s) \geq 1. \label{eq:app_r_and_rs}
\end{equation}

In the next step, we will show that $|\varphi-\varphi_s/2| \leq \pi/2$ which means that $\cos\left(\varphi-\varphi_s/2 \right)$ is nonnegative. It is known fact that the sum of two complex numbers with phases $\delta_1$ and $\delta_2$, where $\delta_1 \leq \delta_2$ and $\delta_1, \delta_2 \in [-\pi,\pi)$ , yields a complex number with the phase $\delta \in [-\pi,\pi)$, that is
\begin{align}
  \delta \in [\delta_1,\delta_2] &\text{ if } |\delta_1-\delta_2| < \pi, \\
  \delta \in [\delta_2,\delta_1] &\text{ if } |\delta_1-\delta_2| > \pi, \\
  \delta = \delta_1 \text{ or } \delta = \delta_2 &\text{ if } |\delta_1-\delta_2| = \pi.
\end{align}
This implies that
\begin{align}
|\delta-\delta_2|\leq \pi \wedge |\delta_1-\delta| \leq \pi. \label{eq:app_delta_diff}
\end{align}

The phase $\varphi_s$ calculated from (\ref{eq:app_zs}) is $\varphi_s = 2\varphi - \theta$, where $ |\theta| \leq \pi$ according to (\ref{eq:app_delta_diff}). Then,
\begin{align}
\left| \varphi -\frac{\varphi_s}{2} \right| = \left| \varphi-\varphi+\frac{1}{2}\theta \right| = \frac{1}{2}|\theta| \leq \frac{1}{2} \pi.
\end{align}
Therefore,
\begin{equation}
  \cos\left(\varphi-\frac{\varphi_s}{2} \right) \geq 0
\end{equation}
and (\ref{eq:app_r2}) gives,
\begin{equation}
  r_2 \geq 1.
\end{equation}
This means that the amplification of $G_2(s)$ for all frequencies is greater or equal to one. Since $G_1(s) = G_2(s)^{-1}$ (\ref{eq:Ginv}), it means that the amplification of $G_1(s)$ on all frequencies is less than or equal to one, hence $||G_1(s)||_{\infty} \leq 1$.

\bibliographystyle{elsart-num-sort}

\bibliography{2013-Wave_based_control}

\begin{thebibliography}{10}


\bibitem{Asmar2004}
N.~H. Asmar, {Partial Differential Equations with Fourier Series and Boundary
  Value Problems}, 2nd ed., Pearson Prentice Hall, New Jersey, 2004.

\bibitem{Bamieh2012b}
B.~Bamieh, M.~R. Jovanovic, P.~Mitra, S.~Patterson, {Coherence in Large-Scale
  Networks: Dimension-Dependent Limitations of Local Feedback}, IEEE
  Transactions on Automatic Control 57~(9) (2012) 2235--2249.

\bibitem{Barooah2005a}
P.~Barooah, J.~Hespanha, {Error amplification and disturbance propagation in
  vehicle strings with decentralized linear control}, in: 44th IEEE Conference
  on Decision and Control, No. theorem 3, 2005, pp. 4964--4969.

\bibitem{Barooah2009}
P.~Barooah, P.~Mehta, J.~Hespanha, {Mistuning-Based Control Design to Improve
  Closed-Loop Stability Margin of Vehicular Platoons}, IEEE Transactions on
  Automatic Control 54~(9) (2009) 2100--2113.

\bibitem{brancik1999}
L.~Brancik, {Programs for fast numerical inversion of Laplace transforms in
  MATLAB language environment}, Conference MATLAB (1999) 27--39.

\bibitem{chu_decentralized_1974}
K.-c. Chu, {Decentralized control of high-speed vehicular strings},
  Transportation Science 8~(4) (1974) 361--384.

\bibitem{Cosgriff1969}
R.~Cosgriff, {The asymptotic approach to traffic dynamics}, IEEE Transactions
  on Systems Science and Cybernetics~(4) (1969) 361--368.

\bibitem{dehoog1982}
F.~R. de~Hoog, J.~H. Knight, A.~N. Stokes, {An Improved Method for Numerical
  Inversion of Laplace Transforms}, SIAM Journal on Scientific and Statistical
  Computing 3~(3) (1982) 357--366.

\bibitem{Eyre1998a}
J.~Eyre, D.~Yanakiev, I.~Kanellakopoulos, {A Simplified Framework for String
  Stability Analysis of Automated Vehicles∗}, Vehicle System Dynamics.

\bibitem{Flotow1985}
A.~H. Flotow, {Disturbance propagation in structural networks}, Journal of
  Sound and Vibration 106~(3) (1986) 433--450.

\bibitem{Flotow1986}
A.~H.~V. Flotow, {Traveling wave control for large spacecraft structures},
  Journal of Guidance Control and Dynamics 9~(4) (1986) 462--468.

\bibitem{french2003}
A.~P. French, {Vibration's and Waves}, M.I.T. introductory physics series, CBS
  Publishers \& Distributors, 2003.

\bibitem{Halevi2005}
Y.~Halevi, {Control of Flexible Structures Governed by the Wave Equation Using
  Infinite Dimensional Transfer Functions}, Journal of Dynamic Systems,
  Measurement, and Control 127~(4) (2005) 579.

\bibitem{Herman2013}
I.~Herman, D.~Martinec, Z.~Hur\'{a}k, M.~\v{S}ebek, {PDdE-based analysis of
  vehicular platoons with spatio-temporal decoupling}, in: Proceedings of 4th
  IFAC Workshop on Distributed Estimation and Control in Networked Systems
  (NecSys), Koblenz, Germany, 2013, pp. 144--151.

\bibitem{jones1984}
W.~B. Jones, W.~J. Thron, {Continued Fractions: Analytic Theory and
  Applications}, Encyclopedia of Mathematics and its Applications, Cambridge
  University Press, 1984.

\bibitem{Jovanovic2005a}
M.~Jovanovic, B.~Bamieh, {On the ill-posedness of certain vehicular platoon
  control problems}, IEEE Transactions on Automatic Control 50~(9) (2005)
  1307--1321.

\bibitem{Lestas2007}
I.~Lestas, G.~Vinnicombe, {Scalability in heterogeneous vehicle platoons}, 2007
  American Control Conference~(M) (2007) 4678--4683.

\bibitem{Levine1966}
W.~Levine, M.~Athans, {On the optimal error regulation of a string of moving
  vehicles}, Automatic Control, IEEE Transactions on.

\bibitem{Lin2011}
F.~Lin, M.~Fardad, M.~R. Jovanovic, {Optimal Control of Vehicular Formations
  With Nearest Neighbor Interactions}, IEEE Transactions on Automatic Control
  57~(9) (2012) 2203--2218.

\bibitem{Martinec2014}
D.~Martinec, I.~Herman, Z.~Hur\'{a}k, M.~\v{S}ebek, {Augmentation of a
  bidirectional platooning controller by wave absorption at the leader}, in:
  (Submitted to the) 3th European Control Conference, 2014.

\bibitem{Martinec2012}
D.~Martinec, M.~Sebek, Z.~Hurak, {Vehicular platooning experiments with racing
  slot cars}, 2012 IEEE International Conference on Control Applications (2012)
  166--171.

\bibitem{Melzer1971}
S.~Melzer, B.~Kuo, {A closed-form solution for the optimal error regulation of
  a string of moving vehicles}, IEEE Transactions on Automatic Control 16~(1)
  (1971) 50--52.

\bibitem{Middleton2010}
R.~H. Middleton, J.~H. Braslavsky, {String Instability in Classes of Linear
  Time Invariant Formation Control With Limited Communication Range}, IEEE
  Transactions on Automatic Control 55~(7) (2010) 1519--1530.

\bibitem{Nagase2005}
K.~Nagase, H.~Ojima, Y.~Hayakawa, {Wave-based Analysis and Wave Control of
  Ladder Networks}, 44th IEEE Conference on Decision and Control (2005)
  5298--5303.

\bibitem{OConnor2006}
W.~J. O’Connor, {Wave-echo control of lumped flexible systems}, Journal of
  Sound and Vibration 298~(4-5) (2006) 1001--1018.

\bibitem{OConnor2007}
W.~J. O'Connor, {Wave-Based Analysis and Control of Lump-Modeled Flexible
  Robots}, IEEE Transactions on Robotics 23~(2) (2007) 342--352.

\bibitem{Ojima2001}
H.~Ojima, K.~Nagase, Y.~Hayakawa, {Wave-based analysis and wave control of
  damped mass-spring systems}, 40th IEEE Conference on Decision and Control
  8~(December) (2001) 2574--2579.

\bibitem{Peled2012}
I.~Peled, W.~O'Connor, Y.~Halevi, {On the relationship between wave based
  control, absolute vibration suppression and input shaping}, Mechanical
  Systems and Signal Processing (2012) 1--11.

\bibitem{Seiler2004a}
P.~Seiler, A.~Pant, K.~Hedrick, {Disturbance Propagation in Vehicle Strings},
  IEEE Transactions on Automatic Control 49~(10) (2004) 1835--1841.

\bibitem{Shaw2007}
E.~Shaw, J.~K. Hedrick, {Controller design for string stable heterogeneous
  vehicle strings}, in: 46th IEEE Conference on Decision and Control, IEEE,
  2007, pp. 2868--2875.

\bibitem{Swaroop1996}
D.~Swaroop, J.~Hedrick, {String stability of interconnected systems}, IEEE
  Transactions on Automatic Control 41~(3) (1996) 349--357.

\bibitem{vaughan_application_1968}
D.~R. Vaughan, {Application of Distributed Parameter Concepts to Dynamic
  Analysis and Control of Bending Vibrations}, Journal of Basic Engineering
  90~(2) (1968) 157--166.

\bibitem{Weeks1966}
W.~Weeks, {Numerical inversion of Laplace transforms using Laguerre functions},
  Journal of the ACM (JACM).

\end{thebibliography}

\end{document}